\documentclass[journal]{IEEEtran}
\newcommand{\papertitle}{Equivalence of Open-Loop and Closed-Loop Operation of SAW Resonators and Delay Lines}

\usepackage{cite}

\ifCLASSINFOpdf
  \usepackage[pdftex]{graphicx}
\else
\fi
\usepackage{amsmath}
\usepackage{amssymb}
\usepackage{sistyle}

\usepackage{tikz}
\usetikzlibrary{arrows.meta}

\usepackage{adjustbox}

\usepackage{hyperref}
\hypersetup{
    unicode=false,
    pdftoolbar=true,
    pdfmenubar=true,
    pdffitwindow=false,
    pdfstartview={FitH},
    pdftitle={\papertitle},
    pdfauthor={Durdaut et al.},
    pdfsubject={\papertitle},
    pdfcreator={Durdaut et al.},
    pdfproducer={Kiel University},
    pdfnewwindow=true,
    colorlinks=true,
    linkcolor=black,
    citecolor=black,
    filecolor=black,
    urlcolor=black
}

\usepackage{graphicx}
\usepackage{caption}
\usepackage{subcaption}

\usepackage{epstopdf}

\usepackage{cancel}

\usepackage{mathtools}

\usepackage{cuted}
\usepackage{mathrsfs}

\usepackage{array}
\begin{document}
\bstctlcite{IEEEexample:BSTcontrol}
%
\title{\papertitle}
%
%
%

\author{Phillip~Durdaut, 
		Michael~H\"oft, 
        Jean-Michel~Friedt,
		and Enrico~Rubiola\\
\thanks{P.~Durdaut and M.~H\"oft are with the Chair of Microwave Engineering, Institute of Electrical Engineering and Information Technology, Faculty of Engineering, Kiel University, Kaiserstr. 2, 24143 Kiel, Germany.}
\thanks{E.~Rubiola and J.-M.~Friedt are with the FEMTO-ST Institute, Department of Time and Frequency, Universit\'{e} de Bourgogne Franche-Comt\'{e} (UBFC), and CNRS, ENSMM, 26 Rue de l'\'{E}pitaphe, 25000 Besan\c{c}on, France. E.~Rubiola is also with the Physics Metrology Division, Istituto Nazionale di Ricerca Metrologica (INRiM), Strada Delle Cacce 91, 10135 Torino, Italy.}}

%
%

\markboth{Durdaut \MakeLowercase{\textit{et al.}}: \papertitle}
{}
%



\maketitle

\begin{abstract}
Surface acoustic wave (SAW) sensors in the form of two-port resonators or delay lines are widely used in various fields of application. Readout of such sensors is achieved by electronic systems operating either in an open-loop or in a closed-loop configuration. The mode of operation of the sensor system is usually chosen based on requirements like e.g. bandwidth, dynamic range, linearity, costs, and immunity against environmental influences. Because the limit of detection (LOD) at the output of a sensor system is often one of the most important figures of merit, both readout structures, i.e. open-loop and closed-loop systems, are analyzed in terms of the minimum achievable LOD. Based on a comprehensive phase noise analysis of these structures for both, resonant sensors and delay line sensors, expressions for the various limits of detection are derived. Under generally valid conditions equivalence of open-loop and closed-loop operation is shown for both types of sensors. These results are not only valid for SAW devices but are also applicable to all kinds of phase sensitive sensors.
\end{abstract}


%
\IEEEpeerreviewmaketitle

%
%
%
%


\section{Introduction}
\label{sec:introduction}
Among many others, surface acoustic wave (SAW) sensors are widely used in various fields of application \cite{White.1985,Liu.2016}. SAW sensors for measuring temperature \cite{Neumeister.1989,Hauden.1981}, pressure \cite{Scherr.1996,Jungwirth.2002}, electric fields \cite{Fransen.1997}, magnetic fields \cite{Smole.2003,Kittmann.2018}, humidity \cite{Caliendo.1993}, and vibration \cite{Filipiak.2011} or for the detection of gases \cite{Devkota.2016} and biorelevant molecules \cite{Laenge.2008,Hirst.2008}, respectively, have been reported.

In this paper, two-port sensors are considered that consist of one input- and one output interdigital transducer (IDT) structured on the piezoelectric substrate to efficiently convert between electrical and mechanic waves \cite{White.1965}. Two SAW device structures are most widely used. A \textit{delay line} essentially consists of two IDTs placed in some distance apart, whereas a \textit{resonator} has additional reflector gratings to confine the wave energy inside a resonant cavity \cite[p. 141]{Fischerauer.2008}. In most cases, the SAW device is coated with a certain material that interacts with the physical quantity to be measured, and in turn, leads to an alteration of the wave propagating along the substrate's surface. Thus, the transceived signals of such coated sensors are generally modulated in its phase and in its amplitude.

For the readout of SAW sensors two structures are most common. A straightforward approach is to compare the sensor's output signal with a local oscillator (LO) signal fed into the sensor in an \textit{open-loop} configuration \cite{Blomley.1973}. Such systems not only allow for the detection of both amplitude and phase changes but are also suited for the characterization of the frequency response of the sensor. However, especially due to the needed LO, these systems are often complex \cite{Rabus.2013,Liu.2017}. In the most common readout structure, the SAW sensor is inserted into the feedback loop of an amplifier, thus forming a closed-loop system in which the oscillating signal is frequency-modulated when the phase response of the sensor changes \cite{Lewis.1974,Parker.1982,Parker.1988}. Such systems appear to be simple \cite{Viens.1990,Schmitt.2001}, but mostly also require a reference oscillator for the frequency detection. In addition, a self-oscillating sensor system is, without introducing additional expense, usually only suitable for the detection of changes in the sensor's phase response because variations in the oscillator signal's amplitude are strongly suppressed by the saturation of the internal amplifier.

In general, depending on the application of a sensor system, properties like e.g. bandwidth, dynamic range, linearity, and immunity against environmental influences are required. However, for high-end sensor systems the limit of detection (LOD) is often the most important figure of merit. In this paper, open-loop and closed-loop sensor readout systems are investigated and compared in terms of the achievable LOD.

This paper is organized as follows: Sec.~\ref{sec:sensitivity} introduces open-loop and closed-loop readout systems for both resonant and delay line sensors. In Sec.~\ref{sec:phase_noise} expressions for describing the phase noise behavior of both readout systems and for both types of sensors are derived. Based on these results, the LOD for the various cases are calculated in Sec.~\ref{sec:limit_of_detection} where the equivalence of the LOD between open-loop and closed-loop systems is shown. This article finishes with an additional consideration of the time domain uncertainty in Sec.~\ref{sec:time_domain_uncertainty} and a summary of the findings in Sec.~\ref{sec:conclusion}.

\section{Sensitivity}
\label{sec:sensitivity}
\subsection{Open-Loop and Closed-Loop Readout Systems}
\label{subsec:sensitivity_open_loop_and_closed_loop_readout_systems}

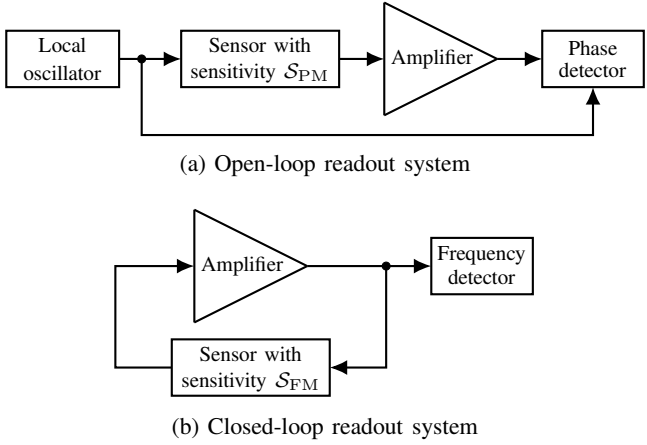
\begin{figure}[t]
	\centering
	\begin{subfigure}[t]{0.5\textwidth}
		\centering
		\trimbox{0cm 0cm 0cm 0cm}{\usetikzlibrary{arrows.meta}

\begin{tikzpicture}[scale=0.75]

\draw[thick]  (-5.5,2) rectangle (-7.5,1);
\node  [align=center] at (-6.5,1.75) {\footnotesize Local};
\node  [align=center] at (-6.5,1.25) {\footnotesize oscillator};

\draw[thick,-{Latex[width=2mm]}] (-5.5,1.5) -- (-4.4,1.5);

\draw[fill=black](-5.1,1.5) circle (2pt);
\draw[thick,-{Latex[width=2mm]}] (-5.1,1.5) -- (-5.1,0.15) -- (2.9,0.15) -- (2.9,1);

\draw[thick]  (-1.6,2) rectangle (-4.4,1);
\node  [align=center] at (-3,1.75) {\footnotesize Sensor with};
\node  [align=center] at (-3,1.25) {\footnotesize sensitivity $\mathcal{S}_{\mathrm{PM}}$};

\draw[thick,-{Latex[width=2mm]}] (-1.6,1.5) -- (-0.8,1.5);

\draw[thick] (-0.8,2.5) -- (-0.8,0.5);
\draw[thick] (1.2,1.5) -- (-0.8,2.5);
\draw[thick] (1.2,1.5) -- (-0.8,0.5);
\node  [align=center] at (0.05,1.5) {\footnotesize Amplifier};

\draw[thick,-{Latex[width=2mm]}] (1.2,1.5) -- (2,1.5);

\draw[thick]  (2,2) rectangle (3.8,1);
\node  [align=center] at (2.9,1.7) {\footnotesize Phase};
\node  [align=center] at (2.9,1.3) {\footnotesize detector};

\end{tikzpicture}}
		\caption{Open-loop readout system}
		\label{fig:readout_open_loop}
	\end{subfigure}
	~
	\begin{subfigure}[t]{0.5\textwidth}
		\centering
		\trimbox{0cm 0cm 0cm -0.4cm}{\usetikzlibrary{arrows.meta}

\begin{tikzpicture}[scale=0.75]

\draw[thick] (-1,2.5) -- (-1,0.5);
\draw[thick] (1,1.5) -- (-1,2.5);
\draw[thick] (1,1.5) -- (-1,0.5);
\node  [align=center] at (-0.15,1.5) {\footnotesize Amplifier};

\draw[thick] (1.4,0.2) rectangle (-1.4,-0.8);
\node  [align=center] at (0,-0.05) {\footnotesize Sensor with};
\node  [align=center] at (0,-0.55) {\footnotesize sensitivity $\mathcal{S}_{\mathrm{FM}}$};

\draw[thick,-{Latex[width=2mm]}] (1,1.5) -- (2.4,1.5) -- (2.4,-0.3) -- (1.4,-0.3);
\draw[thick,-{Latex[width=2mm]}] (-1.4,-0.3) -- (-2.4,-0.3) -- (-2.4,1.5) -- (-1,1.5);
\draw[thick,-{Latex[width=2mm]}] (2.4,1.5) -- (3.2,1.5);
\draw[fill=black] (2.4,1.5) circle (2pt);

\draw[thick] (3.2,2) rectangle (5,1);
\node  [align=center] at (4.1,1.7) {\footnotesize Frequency};
\node  [align=center] at (4.1,1.3) {\footnotesize detector};

\end{tikzpicture}}
		\caption{Closed-loop readout system}
		\label{fig:readout_closed_loop}
	\end{subfigure}
	~
	\caption{Basic structures of open-loop (\subref{fig:readout_open_loop}) and closed-loop (\subref{fig:readout_closed_loop}) sensor readout systems. In open-loop systems the signal transceived by the sensor is phase-modulated with a sensitivity $\mathcal{S}_{\mathrm{PM}}$ whereas the oscillating signal in closed-loop operation is frequency-modulated with a sensitivity $\mathcal{S}_{\mathrm{FM}}$.}
	\label{fig:readout_open_loop_and_closed_loop}
\end{figure}

Fig.~\ref{fig:readout_open_loop_and_closed_loop} depicts the basic structures of an open-loop and a closed-loop sensor readout system. In an open-loop system (Fig.~\ref{fig:readout_open_loop}), a signal derived from an LO is fed into the sensor and afterwards usually amplified. In this kind of system the sensor's transceived signal is phase-modulated with the sensitivity $\mathcal{S}_\mathrm{PM}$. This value is given in units of $\mathrm{rad}/\mathrm{au}$, where $\mathrm{au}$ is an arbitrary unit and depends on the physical quantity to be detected by the sensor. For example, this could be $\mathrm{K}$ for temperature sensors, $\mathrm{Pa}$ for pressure sensors, $\mathrm{T}$ for magnetic field sensors, $\mathrm{A}$ for current sensors, $\mathrm{m}$ for distance sensors, etc. To reconstruct the modulation signal a phase detector, e.g. a mixer, is utilized. For coherent phase detectors the phase noise of the LO is usually negligible because it is largely suppressed \cite{Durdaut.2018}.

The closed-loop readout system (Fig.~\ref{fig:readout_closed_loop}) oscillates when the amplifier's gain is large enough to compensate for the sensor's losses (loop gain $>1$) and when constructive superposition of the periodic signal with a certain frequency (loop phase equal to $2 \pi n, n \in \mathbb{N}_0$) is assured. These conditions are known as the \textit{Barkhausen stability criterion} \cite{Barkhausen.1935}. Since the magnitude frequency response of common sensors show a more or less strong dependence on the frequency, i.e. a certain bandwidth, the amplitude condition is usually only fulfilled for one frequency (resonator with small bandwidth) or for a small number of frequencies (delay lines with relatively high bandwidths). If the phase response of the frequency determining element, i.e. the sensor, slightly changes due to an extrinsic influence, technically speaking the loop phase criterion is no longer fulfilled such that the oscillating frequency changes. Thus, the oscillating signal is frequency-modulated by the externally changing physical quantity to be measured with the sensitivity $\mathcal{S}_\mathrm{FM}$ given in units of $\mathrm{Hz}/\mathrm{au}$. To reconstruct the modulation signal a frequency detector, e.g. a phase-locked loop (PLL) \cite{Hsieh.1996}, is commonly utilized.

\begin{figure}[t]
	\centering
	\begin{subfigure}[t]{0.23\textwidth}
		\centering
		\trimbox{0.5cm 0cm 0cm 0cm}{\usetikzlibrary{arrows.meta}

\begin{tikzpicture}

\draw[thin, densely dotted] (-0.2,-3) -- (-0.2,-1);
\draw[thin, densely dotted] (0.2,-3) -- (0.2,-1);
\draw[thick,-{Latex[width=2mm,length=1mm]}] (-0.2,-1.2) -- (0.2,-1.2);
\draw[thick,-{Latex[width=2mm,length=1mm]}] (0.2,-1.2) -- (-0.2,-1.2);
\node  [align=left] at (1.1,-1) {Frequency};
\node  [align=left] at (1.1,-1.4) {detuning};

\draw[thick,-{Latex[width=2mm]}] (-1.6,0) -- (1.8,0);
\draw[thick,-{Latex[width=2mm]}] (-1.6,0) -- (-1.6,2);
\draw[thick] (0,-0.05) -- (0,0.05);

\node  [align=center] at (-1.6,2.25) {$|H_{\mathrm{R}}(f)|$};
\node  [align=center] at (2.05,0) {$f$};
\node  [align=center] at (0,-0.32) {$f_{\mathrm{R}}$};

\node (v1a) at (-1.5,0) {};
\node (v2a) at (-1,0.2) {};
\node (v3a) at (-0.57,0.8) {};
\node (v4a) at (-0.2,1.6) {};
\node (v5a) at (0.17,0.8) {};
\node (v6a) at (0.6,0.2) {};
\node (v7a) at (1.1,0) {};
\draw[thin,densely dashed] plot[smooth, tension=.7] coordinates {(v1a) (v2a) (v3a) (v4a) (v5a) (v6a) (v7a)};

\node (v1) at (-1.3,0) {};
\node (v2) at (-0.8,0.2) {};
\node (v3) at (-0.37,0.8) {};
\node (v4) at (0,1.6) {};
\node (v5) at (0.37,0.8) {};
\node (v6) at (0.8,0.2) {};
\node (v7) at (1.3,0) {};
\draw[thick] plot[smooth, tension=.7] coordinates {(v1) (v2) (v3) (v4) (v5) (v6) (v7)};

\node (v1b) at (-1.1,0) {};
\node (v2b) at (-0.6,0.2) {};
\node (v3b) at (-0.17,0.8) {};
\node (v4b) at (0.2,1.6) {};
\node (v5b) at (0.57,0.8) {};
\node (v6b) at (1,0.2) {};
\node (v7b) at (1.5,0) {};
\draw[thin,densely dashed] plot[smooth, tension=.7] coordinates {(v1b) (v2b) (v3b) (v4b) (v5b) (v6b) (v7b)};

\draw[thick,-{Latex[width=2mm]}] (-1.6,-3) -- (1.8,-3);
\draw[thick,-{Latex[width=2mm]}] (-1.6,-3) -- (-1.6,-1);
\draw[thick] (0,-0.05-3) -- (0,0.05-3);

\node  [align=center] at (-1.6,2.25-3) {$\varphi_{\mathrm{R}}(f)$};
\node  [align=center] at (2.05,-3) {$f$};
\node  [align=center] at (0,-3.32) {$f_{\mathrm{R}}$};

\node (v1a) at (-1.5,-1.4) {};
\node (v2a) at (-0.8,-1.5) {};
\node (v3a) at (-0.4,-1.9) {};
\node (v4a) at (-0.2,-2.2) {};
\node (v5a) at (0,-2.5) {};
\node (v6a) at (0.4,-2.9) {};
\node (v7a) at (1.1,-3) {};
\draw[thin,densely dashed] plot[smooth, tension=.5] coordinates {(v1a) (v2a) (v3a) (v4a) (v5a) (v6a) (v7a)};

\node (v1) at (-1.3,-1.4) {};
\node (v2) at (-0.6,-1.5) {};
\node (v3) at (-0.2,-1.9) {};
\node (v4) at (0,-2.2) {};
\node (v5) at (0.2,-2.5) {};
\node (v6) at (0.6,-2.9) {};
\node (v7) at (1.3,-3) {};
\draw[thick] plot[smooth, tension=.5] coordinates {(v1) (v2) (v3) (v4) (v5) (v6) (v7)};

\node (v1b) at (-1.1,-1.4) {};
\node (v2b) at (-0.4,-1.5) {};
\node (v3b) at (0,-1.9) {};
\node (v4b) at (0.2,-2.2) {};
\node (v5b) at (0.4,-2.5) {};
\node (v6b) at (0.8,-2.9) {};
\node (v7b) at (1.5,-3) {};
\draw[thin,densely dashed] plot[smooth, tension=.5] coordinates {(v1b) (v2b) (v3b) (v4b) (v5b) (v6b) (v7b)};

\end{tikzpicture}}
		\caption{Resonant sensor}
		\label{fig:resonator_magnitude_phase}
	\end{subfigure}
	~
	\begin{subfigure}[t]{0.23\textwidth}
		\centering
		\trimbox{0.5cm 0cm 0cm 0cm}{\usetikzlibrary{arrows.meta}

\begin{tikzpicture}

\draw[thin, densely dotted] (1.3,-1.8) -- (-1.3,-1.8);
\draw[thin, densely dotted] (1.3,-2.2) -- (-1.3,-2.2);
\draw[thick,-{Latex[width=2mm,length=1mm]}] (1.2,-2.2) -- (1.2,-1.8);
\draw[thick,-{Latex[width=2mm,length=1mm]}] (1.2,-1.8) -- (1.2,-2.2);
\node  [align=center] at (1.2,-1.1) {Phase};
\node  [align=center] at (1.2,-1.5) {detuning};

\draw[thick,-{Latex[width=2mm]}] (-1.6,0) -- (1.8,0);
\draw[thick,-{Latex[width=2mm]}] (-1.6,0) -- (-1.6,2);
\draw[thick] (0,-0.05) -- (0,0.05);

\node  [align=center] at (-1.6,2.25) {$|H_{\mathrm{D}}(f)|$};
\node  [align=center] at (2.05,0) {$f$};
\node  [align=center] at (0,-0.32) {$f_{\mathrm{D}}$};

\node (v1a) at (-1.3,0) {};
\node (v2a) at (-1.3,0.2) {};
\node (v3a) at (-1.22,0.8) {};
\node (v4aa) at (-1,1.7) {};
\node (v4ab) at (1,1.7) {};
\node (v5a) at (1.22,0.8) {};
\node (v6a) at (1.3,0.2) {};
\node (v7a) at (1.3,0) {};
\draw[thin,densely dashed] plot[smooth, tension=.2] coordinates {(v1a) (v2a) (v3a) (v4aa) (v4ab) (v5a) (v6a) (v7a)};

\node (v1) at (-1.3,0) {};
\node (v2) at (-1.3,0.2) {};
\node (v3) at (-1.22,0.8) {};
\node (v4a) at (-1,1.6) {};
\node (v4b) at (1,1.6) {};
\node (v5) at (1.22,0.8) {};
\node (v6) at (1.3,0.2) {};
\node (v7) at (1.3,0) {};
\draw[thick] plot[smooth, tension=.2] coordinates {(v1) (v2) (v3) (v4a) (v4b) (v5) (v6) (v7)};

\node (v1b) at (-1.3,0) {};
\node (v2b) at (-1.3,0.2) {};
\node (v3b) at (-1.22,0.8) {};
\node (v4ba) at (-1,1.5) {};
\node (v4bb) at (1,1.5) {};
\node (v5b) at (1.22,0.8) {};
\node (v6b) at (1.3,0.2) {};
\node (v7b) at (1.3,0) {};
\draw[thin,densely dashed] plot[smooth, tension=.2] coordinates {(v1b) (v2b) (v3b) (v4ba) (v4bb) (v5b) (v6b) (v7b)};

\draw[thick,-{Latex[width=2mm]}] (-1.6,-3) -- (1.8,-3);
\draw[thick,-{Latex[width=2mm]}] (-1.6,-3) -- (-1.6,-1);
\draw[thick] (0,-0.05-3) -- (0,0.05-3);

\node  [align=center] at (-1.6,2.25-3) {$\varphi_{\mathrm{D}}(f)$};
\node  [align=center] at (2.05,-3) {$f$};
\node  [align=center] at (0,-3.32) {$f_{\mathrm{D}}$};

\draw[thin,densely dashed] (-1,-1.4) -- (1,-2.2);
\draw[thick,-] (-1,-1.4) -- (1,-2.6);
\draw[thin,densely dashed] (-1,-1.4) -- (1,-3);

\end{tikzpicture}}
		\caption{Delay line sensor}
		\label{fig:delay_line_magnitude_phase}
	\end{subfigure}
	~
	\caption{General impact of a change in the physical quantity to be detected on the frequency response of a resonant sensor (\subref{fig:resonator_magnitude_phase}) and on a delay line sensor (\subref{fig:delay_line_magnitude_phase}). Changes in magnitude are not discussed in this article because of amplitude compression in closed-loop readout systems.}
	\label{fig:resonator_and_delay_line_magnitude_phase}
\end{figure}
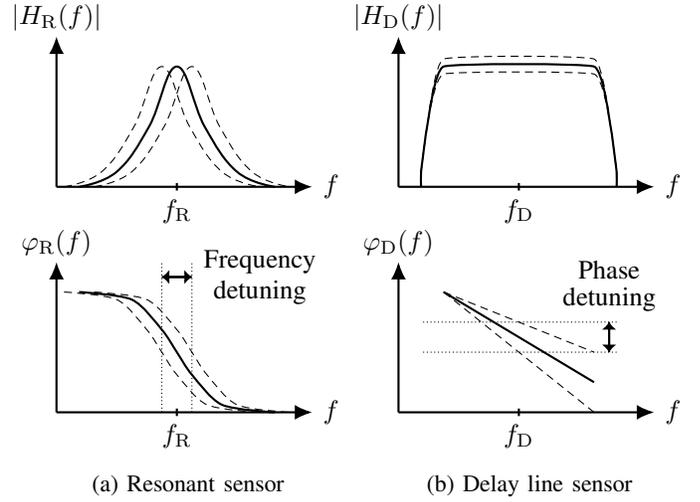

\begin{table}[!t]
\renewcommand{\arraystretch}{1.5}
\caption{Relations between open-loop and closed-loop sensitivities for resonant sensors and delay line sensors.}
\label{tab:relations_between_sensitivities}
\centering
\begin{tabular}{l|c|c|}
\cline{2-3}
                        & Open-loop sensitivity & Closed-loop sensitivity \\
                        & $\mathcal{S}_{\mathrm{PM}}$ [$\mathrm{rad}/\mathrm{au}$] & $\mathcal{S}_{\mathrm{FM}}$ [$\mathrm{Hz}/\mathrm{au}$] \\ \hline
\multicolumn{1}{|l|}{Resonant sensor} & $-2 \pi \tau_{\mathrm{R}}~\mathcal{S}_{\mathrm{R}}$ & $\mathcal{S}_{\mathrm{R}}$ \\ \hline
\multicolumn{1}{|l|}{Delay line sensor} & $\mathcal{S}_{\mathrm{D}}$ & $-1/(2 \pi \tau_{\mathrm{D}})~\mathcal{S}_{\mathrm{D}}$ \\ \hline
\end{tabular}
\end{table}

\subsection{Resonator}
\label{subsec:sensitivity_resonator}

According to the physical principle of a resonant sensor with a quality factor $Q$, a natural frequency $f_{\mathrm{R}}$, and a \SI{-3}{dB} bandwidth of ${B_{_\mathrm{R}} = f_{\mathrm{R}}/Q}$, its sensitivity $\mathcal{S}_\mathrm{R}$ is given in units of ${\mathrm{Hz}/\mathrm{au}}$. Thus, it holds that
\begin{equation}
	\mathcal{S}_\mathrm{FM} = \mathcal{S}_\mathrm{R}
	\label{eqn:SFMresonator}
\end{equation}
when utilizing a resonant sensor in a closed-loop readout system. Throughout this paper, it is assumed that the considered resonator has a low damping factor (i.e. a large quality factor) and can be described by a second-order differential equation. For such a resonator with the frequency response ${H_{\mathrm{R}}(f) = |H_{\mathrm{R}}| \exp(-j \varphi_{\mathrm{R}}(f))}$ (see Fig.~\ref{fig:resonator_magnitude_phase}) the slope of the linear phase response in the vicinity of the natural frequency is given by \cite[p. 71 f.]{Rubiola.2009}
\begin{equation}
	\frac{\mathrm{d} \varphi_{\mathrm{R}}(f)}{\mathrm{d} f} = -\frac{2 Q}{f_{\mathrm{R}}}.
\end{equation}
Thus, detuning the sensor's resonance frequency by $\mathcal{S}_\mathrm{R}$ results in phase changes (i.e. in an open-loop sensitivity) of 
\begin{equation}
	\mathcal{S}_{\mathrm{PM}} = \frac{\mathrm{d} \varphi_{\mathrm{R}}(f)}{\mathrm{d} f} \mathcal{S}_{\mathrm{R}} = -\frac{2 Q}{f_{\mathrm{R}}} \mathcal{S}_{\mathrm{R}} = -2 \pi \tau_{\mathrm{R}} \mathcal{S}_{\mathrm{R}}
	\label{eqn:SPMresonator}
\end{equation}
where $\tau_{\mathrm{R}} = Q/(\pi f_{\mathrm{R}})$ is the resonator's relaxation time.

\subsection{Delay Line}
\label{subsec:sensitivity_delay_line}

For delay line sensors, the phase of the transceived signal is altered by the physical quantity to be detected. Therefore, the sensitivity of a delay line sensor $\mathcal{S}_\mathrm{D}$ is given in units of $\mathrm{rad}/\mathrm{au}$, and is thus equal to the open-loop sensitivity
\begin{equation}
	\mathcal{S}_\mathrm{PM} = \mathcal{S}_\mathrm{D}.
	\label{eqn:SPMdelayline}
\end{equation}
Delay line sensors are characterized by their center frequency $f_{\mathrm{D}}$ and their time delay
\begin{equation}
	\tau_{\mathrm{D}} = -\frac{1}{2 \pi} \frac{\mathrm{d} \varphi_{\mathrm{D}}(f)}{\mathrm{d} f}
\end{equation}
where $-\mathrm{d} \varphi_{\mathrm{D}}(f)/\mathrm{d} f$ is the slope of the linear phase response in the delay line sensor's passband with a bandpass characteristic and a \SI{-3}{dB} bandwidth denoted to as $B_{\mathrm{D}}$. The frequency response of a delay line sensor ${H_{\mathrm{D}}(f) = |H_{\mathrm{D}}| \exp(-j \varphi_{\mathrm{D}}(f))}$ is depicted in Fig.~\ref{fig:delay_line_magnitude_phase}. If the phase response of the sensor changes with $\mathcal{S}_\mathrm{D}$ the closed-loop sensitivity yields
\begin{equation}
	\mathcal{S}_{\mathrm{FM}} = \frac{\mathrm{d} f}{\mathrm{d} \varphi_{\mathrm{D}}(f)} \mathcal{S}_{\mathrm{D}} = -\frac{1}{2 \pi \tau_{\mathrm{D}}} \mathcal{S}_{\mathrm{D}}.
	\label{eqn:SFMdelayline}
\end{equation}

Tab.~\ref{tab:relations_between_sensitivities} summarizes the relations between sensitivities in open-loop and closed-loop readout systems for resonant sensors and delay line sensors.

\section{Phase Noise}
\label{sec:phase_noise}
Assuming an arbitrary signal $x(t)$ that describes the physical quantity to be measured in units of $\mathrm{au}$, the phase-modulated signal in an open-loop system can be expressed as
\begin{align}
	s_{\mathrm{PM}}(t) \propto \cos\Big( 2 \pi f_0 t + \mathcal{S}_{\mathrm{PM}} x(t) + \psi_{\mathrm{OL}}(t)\Big)
	\label{eqn:sPMt}
\end{align}
where the carrier signal with the frequency $f_0$ is impaired by random phase fluctuations $\psi_{\mathrm{OL}}(t)$ in units of $\mathrm{rad}$ due to phase noise introduced by the readout electronics and by the sensor itself. The frequency modulated signal in a closed-loop system is also impaired by random phase fluctuations $\psi_{\mathrm{CL}}(t)$ in units of $\mathrm{rad}$
\begin{align}
	s_{\mathrm{FM}}(t) \propto \cos\bigg( 2 \pi f_0 t + 2 \pi \mathcal{S}_{\mathrm{FM}} \int_0^{t}{x(\tilde{t})~\mathrm{d}\tilde{t}} + \psi_{\mathrm{CL}}(t)\bigg)
	\label{eqn:sFMtphasenoise}
\end{align}
which can, alternatively, also be described by random frequency fluctuations $f_{\mathrm{CL}}(t)$ in units of $\mathrm{Hz}$
\begin{align}
	s_{\mathrm{FM}}(t) \propto \cos\bigg( 2 \pi f_0 t + 2 \pi \Big( \mathcal{S}_{\mathrm{FM}} \int_0^{t}{x(\tilde{t})~\mathrm{d}\tilde{t}}\notag\\
	&\hspace{-4.5cm} + \int_0^{t}{f_{\mathrm{CL}}(\tilde{t})~\mathrm{d}\tilde{t}} \Big) \bigg).
	\label{eqn:sFMtfrequencynoise}
\end{align}
With the instantaneous frequency being the time derivative of the phase, the relation between random phase fluctuations and random frequency fluctuations in the time domain \cite{Rutman.1978} is given by
\begin{align}
	f_{\mathrm{CL}}(t) = \frac{1}{2 \pi} \frac{\mathrm{d} \psi_\mathrm{CL}(t)}{\mathrm{d} t}.
	\label{eqn:fCLvonpsiCL}
\end{align}

In general, arbitrary random phase fluctuations $\varphi(t)$ are best described by the one-sided power spectral density $S_{\varphi}(f)$ of the random phase fluctuations. An equivalent and widely used representation is $\mathscr{L}(f)$ which is defined as ${\mathscr{L}(f) = 1/2~S_{\varphi}(f)}$ \cite{IEEE.2009}. However, $S_{\varphi}(f)$ is used throughout this paper because it is given in SI units of $\mathrm{rad}^2/\mathrm{Hz}$, and thus makes further conversions more straightforward. A model that has been found useful in describing the frequency dependence of a power spectral density of random phase fluctuations is the power law 
\begin{align}
	S_{\varphi}(f) = \sum_{i=-n}^{0}{b_{i}f^{i}}
	\label{eqn:Sphi_powerlaw}
\end{align}
with usually $n \le 4$. $i = 0$ and $i = -1$ refer to white phase noise and $1/f$ flicker phase noise, respectively, which are the main processes in two-port components \cite[p. 23]{Rubiola.2009} \cite{Boudot.2012}. As will be shown further below, in closed-loop systems white phase noise results in white frequency noise ($i = -2$) and flicker phase noise results in flicker frequency noise ($i = -3$). Higher order effects like random walk of frequency ($i = -4$) are related to environmental changes like e.g. temperature drifts, humidity, and vibrations \cite{Rutman.1978}.

The term ${b_0 f^0 = F k_{\mathrm{B}} T_0/P_0}$ quantifies the constant, i.e. white, phase noise floor where $F$ is the noise figure and ${k_{\mathrm{B}} T_0}$ is the thermal energy. This type of noise is \textit{additive}, which means that $F$ does not change when a carrier signal with power $P_0$ is injected into the according component. When e.g. a sensor and an amplifier are cascaded, the overall phase noise at the output depends on the individual gains and can be calculated by an adaption of the well-known \textit{Friis formula} \cite[p. 49]{Rubiola.2009} \cite{Boudot.2012}. Flicker phase noise is always present, described by the term ${b_{-1}f^{-1}}$. It is a form of \textit{parametric} noise because the carrier is modulated by a near-DC flicker process. Experiments show that $b_{-1}$ is about independent of carrier power $P_0$, thus, the Friis formula does not apply for cascaded two-port components showing flicker phase noise. Instead, the flicker phase noise, i.e. the coefficients $b_{-1}$ of the individual components just adds up \cite{Boudot.2012}.

Phase modulation is difficult to model. Therefore, we transform the radio frequency (RF) schemes into their \textit{phase-space} equivalent, which is a \textit{linear representation} where the signal is the phase of the original RF circuit. This transformation is shown for the open-loop system in Fig.~\ref{fig:open_loop_rf_and_phase} and for the closed-loop system in Fig.~\ref{fig:oscillator_rf_and_phase}, respectively, and extensively discussed later. It is assumed that the gain $A$ of the amplifiers is constant in the frequency range around the sensor's center frequency. Thus, in the phase-space representation an amplifier simply repeats the input phase to its output and has a gain exactly equal to one \cite{Rubiola.2007}. For the frequency dependent transfer function of the sensor in the Laplace domain $H_{\mathrm{S}}(s)$ with the complex angular frequency ${s = \sigma + j \omega}$, the equivalent phase-space representation ${\mathcal{H}_{\mathrm{S}}(s)}$ is calculated with the \textit{phase-step method} \cite[p.~103~ff.]{Rubiola.2009} \cite[Sec.~4]{Rubiola.2010}. This method is based on the well-known property of linear time-invariant (LTI) systems for which the impulse response is the derivative of the step response and the system's transfer function is the Laplace transform of the impulse response. Thus, the phase-space representation of the sensor's transfer function 
\begin{align}
	\mathcal{H}_{\mathrm{S}}(s) = \mathcal{L}\left(\frac{\mathrm{d} h_{\mathrm{S}}(t)}{\mathrm{d}t}\right)
\end{align}
is the Laplace transform of the derivative of the phase step response ${h_{\mathrm{S}}}(t)$ which follows as part of the output signal ${\cos\left( 2 \pi f_0 t + h_{\mathrm{S}}(t) \right)}$ when, in turn, a phase step ${\kappa u(t)}$ with ${\kappa \rightarrow 0}$ as part of the input signal ${\cos\left( 2 \pi f_0 t + \kappa u(t) \right)}$ is fed into the sensor. With ${\kappa \rightarrow 0}$ linearization is obtained which is physically correct for phase noise being usually very small. The term $u(t)$ is the unit-step function also referred to as \textit{Heaviside function}.

For a resonant sensor with the angular natural frequency $\omega_{\mathrm{R}} = 2 \pi f_{\mathrm{R}}$ which can be described by the general transfer function
\begin{align}
	H_{\mathrm{R}}(s) = \frac{\omega_{\mathrm{R}} s}{Q s^2 + \omega_{\mathrm{R}} s + Q \omega_{\mathrm{R}}^2}
\end{align}
the phase-space representative is given by
\begin{equation}
	\mathcal{H}_{\mathrm{R}}(s) = \frac{\pi f_{\mathrm{R}}}{s Q + \pi f_{\mathrm{R}}}.
	\label{eqn:HRs}
\end{equation}
The according magnitude-squared transfer function yields
\begin{equation}
	|\mathcal{H}_{\mathrm{R}}(f)|^2 = \frac{1}{1 + \left(\frac{2 Q f}{f_{\mathrm{R}}}\right)^2}.
	\label{eqn:HRfmagsq}
\end{equation}

The magnitude frequency response of a SAW delay line device is occasionally described using a \textit{sinc} function ${\mathrm{sinc}((f-f_{\mathrm{D}}) / B_{\mathrm{D}} \alpha)}$ where $\alpha$ is a correction factor \cite[p. 80]{Campbell.1998}. Such a function properly can take into account the steepness of the bandpass characteristic and transmission zeros. However, because SAW sensors are always operated in their passband, calculations in this paper are simplified by choosing the transfer function of a bandpass filter to describe the sensor. With the angular center frequency $\omega_{\mathrm{D}} = 2 \pi f_{\mathrm{D}}$ the SAW delay line sensor's frequency response then yields
\begin{align}
	H_{\mathrm{D}}(s) &= \left| \frac{\omega_{\mathrm{D}} B_{\mathrm{D}} s}{f_{\mathrm{D}} s^2 + \omega_{\mathrm{D}} B_{\mathrm{D}} s + f_{\mathrm{D}} \omega_{\mathrm{D}}^2} \right| \cdot e^{-s \tau_{\mathrm{D}}}\notag\\
	&= \frac{1}{\sqrt{1 + \frac{1}{(B_{\mathrm{D}} f)^2} \left( f^2 - f_{\mathrm{D}}^2 \right)^2}} \cdot e^{-s \tau_{\mathrm{D}}}
\end{align}
which results in a phase-space representative given by
\begin{align}
	\mathcal{H}_{\mathrm{D}}(s) &= \left| \frac{\pi B_{\mathrm{D}}}{s + \pi B_{\mathrm{D}}} \right| \cdot e^{-s \tau_{\mathrm{D}}}\notag\\
	&= \frac{1}{\sqrt{1 + \left( \frac{2 f}{B_{\mathrm{D}}} \right)^2}} \cdot e^{-s \tau_{\mathrm{D}}}.
	\label{eqn:HDs}
\end{align}
The according magnitude-squared transfer function yields
\begin{align}
	|\mathcal{H}_{\mathrm{D}}(f)|^2 = \frac{1}{1 + \left(\frac{2 f}{B_{\mathrm{D}}}\right)^2}.
	\label{eqn:HDfmagsq}
\end{align}

\subsection{Phase Noise in Open-Loop Readout System}
\label{subsec:phase_noise_phase_noise_in_open_loop_readout_system}

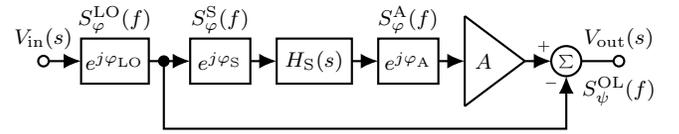
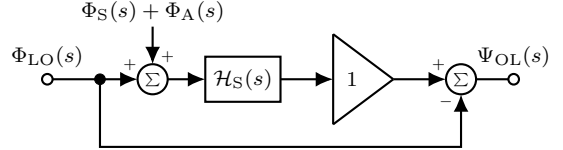
\begin{figure}[t]
	\centering
	\begin{subfigure}[t]{0.5\textwidth}
		\centering
		\trimbox{0cm 0cm 0cm 0cm}{\usetikzlibrary{arrows.meta}

\begin{tikzpicture}

\draw[thick,-{Latex[width=2mm]}] (-6.8,1.5) -- (-6.3,1.5);

\draw[thick,fill=white] (-6.8,1.5) circle (2pt);
\node  [align=center] at (-6.8,1.8) {\footnotesize $V_{\mathrm{in}}(s)$};

\draw[thick]  (-5.4,1.8) rectangle (-6.3,1.2);
\node  [align=center] at (-5.85,1.5) {\footnotesize $e^{j \varphi_\mathrm{LO}}$};
\node  [align=center] at (-5.85,2.05) {\footnotesize$S_{\varphi}^{\mathrm{LO}}(f)$};

\draw[thick,-{Latex[width=2mm]}] (-5.4,1.5) -- (-4.85,1.5);

\draw[thick]  (-4.05,1.8) rectangle (-4.85,1.2);
\node  [align=center] at (-4.45,1.5) {\footnotesize $e^{j \varphi_\mathrm{S}}$};
\node  [align=center] at (-4.45,2.05) {\footnotesize $S_{\varphi}^{\mathrm{S}}(f)$};

\draw[thick,-{Latex[width=2mm]}] (-4.05,1.5) -- (-3.7,1.5);

\draw[thick]  (-2.7,1.8) rectangle (-3.7,1.2);
\node  [align=center] at (-3.2,1.5) {\footnotesize $H_{\mathrm{S}}(s)$};

\draw[thick,-{Latex[width=2mm]}] (-2.7,1.5) -- (-2.35,1.5);

\draw[thick]  (-1.55,1.8) rectangle (-2.35,1.2);
\node  [align=center] at (-1.95,1.5) {\footnotesize $e^{j \varphi_\mathrm{A}}$};
\node  [align=center] at (-1.95,2.05) {\footnotesize $S_{\varphi}^{\mathrm{A}}(f)$};

\draw[thick,-{Latex[width=2mm]}] (-1.55,1.5) -- (-1.2,1.5);

\draw[thick] (-1.2,2.1) -- (-1.2,0.9);
\draw[thick] (-0.4,1.5) -- (-1.2,2.1);
\draw[thick] (-0.4,1.5) -- (-1.2,0.9);
\node  [align=center] at (-0.95,1.5) {\footnotesize $A$};

\draw[thick,-{Latex[width=2mm]}] (-0.4,1.5) -- (-0.05,1.5);

\draw[thick,fill=white] (0.15,1.5) circle (0.2cm);
\node  [align=center] at (0.15,1.5) {\tiny $\sum$};
\node  [align=center] at (-0.15,1.7) {\tiny $+$};
\node  [align=center] at (-0.05,1.2) {\tiny $-$};

\draw[thick,-] (0.35,1.5) -- (0.85,1.5);

\draw[thick,fill=white] (0.85,1.5) circle (2pt);
\node  [align=center] at (0.85,1.8) {\footnotesize $V_{\mathrm{out}}(s)$};
\node  [align=center] at (0.85,1.1) {\footnotesize $S_{\psi}^{\mathrm{OL}}(f)$};

\draw[fill=black](-5.2,1.5) circle (2pt);
\draw[thick,-{Latex[width=2mm]}] (-5.2,1.5) -- (-5.2,0.6) -- (0.15,0.6) -- (0.15,1.3);

\end{tikzpicture}}
		\caption{Open-loop system with random phase contributions}
		\label{fig:open_loop_rf}
	\end{subfigure}
	~
	\begin{subfigure}[t]{0.5\textwidth}
		\centering
		\trimbox{0cm 0cm 0cm -0.4cm}{\usetikzlibrary{arrows.meta}

\begin{tikzpicture}

\draw[thick,-{Latex[width=2mm]}] (-5.9,1.5) -- (-4.7,1.5);

\draw[thick,fill=white] (-5.9,1.5) circle (2pt);
\node  [align=center] at (-5.9,1.8) {\footnotesize $\Phi_{\mathrm{LO}}(s)$};

\draw[thick,-{Latex[width=2mm]}] (-4.3,1.5) -- (-3.8,1.5);

\draw[thick,-{Latex[width=2mm]}] (-4.5,2.2) -- (-4.5,1.7);
\draw[thick,fill=white] (-4.5,1.5) circle (0.2cm);
\node  [align=center] at (-4.5,1.5) {\tiny $\sum$};
\node  [align=center] at (-4.8,1.7) {\tiny $+$};
\node  [align=center] at (-4.3,1.8) {\tiny $+$};
\node  [align=center] at (-4.5,2.4) {\footnotesize $\Phi_{\mathrm{S}}(s) + \Phi_{\mathrm{A}}(s)$};

\draw[thick]  (-2.8,1.8) rectangle (-3.8,1.2);
\node  [align=center] at (-3.3,1.5) {\footnotesize $\mathcal{H}_{\mathrm{S}}(s)$};

\draw[thick,-{Latex[width=2mm]}] (-2.8,1.5) -- (-2.1,1.5);

\draw[thick] (-2.1,2.1) -- (-2.1,0.9);
\draw[thick] (-1.3,1.5) -- (-2.1,2.1);
\draw[thick] (-1.3,1.5) -- (-2.1,0.9);
\node  [align=center] at (-1.85,1.5) {\footnotesize $1$};

\draw[thick,-{Latex[width=2mm]}] (-1.3,1.5) -- (-0.6,1.5);

\draw[thick,fill=white] (-0.4,1.5) circle (0.2cm);
\node  [align=center] at (-0.4,1.5) {\tiny $\sum$};
\node  [align=center] at (-0.7,1.7) {\tiny $+$};
\node  [align=center] at (-0.6,1.2) {\tiny $-$};

\draw[thick,-] (-0.2,1.5) -- (0.3,1.5);

\draw[thick,fill=white] (0.3,1.5) circle (2pt);
\node  [align=center] at (0.3,1.8) {\footnotesize $\Psi_{\mathrm{OL}}(s)$};

\draw[fill=black](-5.2,1.5) circle (2pt);
\draw[thick,-{Latex[width=2mm]}] (-5.2,1.5) -- (-5.2,0.6) -- (-0.4,0.6) -- (-0.4,1.3);

\end{tikzpicture}}
		\caption{Phase-space equivalent system}
		\label{fig:open_loop_phase}
	\end{subfigure}
	~
	\caption{Basic structure of an open-loop sensor readout system together with the random phase contributions of the individual components (\subref{fig:open_loop_rf}). The use of the phase-space equivalent system (\subref{fig:open_loop_phase}) simplifies phase noise analysis as phase noise turns into additive noise but requires the determination of the phase-space equivalent transfer function of the sensor ${\mathcal{H}_{\mathrm{S}}(s)}$.}
	\label{fig:open_loop_rf_and_phase}
\end{figure}

Fig.~\ref{fig:open_loop_rf} depicts the open-loop readout system in the RF domain together with the random phase fluctuations of the input voltage, i.e. the LO, $\varphi_{\mathrm{LO}}$, the sensor $\varphi_{\mathrm{S}}$, and the amplifier $\varphi_{\mathrm{A}}$. The related power spectral densities are denoted by ${S_{\varphi}^{\mathrm{LO}}(f)}$, ${S_{\varphi}^{\mathrm{S}}(f)}$, and ${S_{\varphi}^{\mathrm{A}}(f)}$. As described above, the phase-space representation of the system (Fig.~\ref{fig:open_loop_phase}) is more suited to calculate the overall phase noise at the output of the open-loop system ${S_{\psi}^{\mathrm{OL}}(f)}$. Due to linearity, the Laplace transforms of the phase noise of the sensor ${\Phi_{\mathrm{S}}(s) = \mathcal{L}(\varphi_{\mathrm{S}})}$ and the amplifier ${\Phi_{\mathrm{A}}(s) = \mathcal{L}(\varphi_{\mathrm{A}})}$ can be arranged in front of the phase-equivalent sensor ${\mathcal{H}_{\mathrm{S}}(s)}$. Thus, the phase noise transfer function for both the sensor and the amplifier to the output of the system
\begin{equation}
	\frac{\Psi_{\mathrm{OL}}(s)}{\Phi_{\mathrm{S}}(s)} = \frac{\Psi_{\mathrm{OL}}(s)}{\Phi_{\mathrm{A}}(s)} = \mathcal{H}_{\mathrm{S}}(s)
	\label{eqn:Hss}
\end{equation}
is equal to ${\mathcal{H}_{\mathrm{S}}(s)}$ where ${\Psi_{\mathrm{OL}}(s) = \mathcal{L}(\psi_{\mathrm{OL}})}$. For the phase noise of the LO ${\Phi_{\mathrm{LO}}(s) = \mathcal{L}(\varphi_{\mathrm{LO}})}$ the phase noise transfer function to the output of the open-loop system is given by 
\begin{equation}
	\frac{\Psi_{\mathrm{OL}}(s)}{\Phi_{\mathrm{LO}}(s)} = \mathcal{H}_{\mathrm{S}}(s) - 1.
	\label{eqn:Hssminuseins}
\end{equation}
Thus, the overall power spectral density of the random phase fluctuations at the output of the open-loop sensor system as a function of both the phase noise of the individual components and the frequency response of the sensor yields
\begin{align}
	S_{\psi}^{\mathrm{OL}}(f) = |\mathcal{H}_{\mathrm{S}}(f)|^2~\left(S_{\varphi}^{\mathrm{S}}(f) + S_{\varphi}^{\mathrm{A}}(f)\right)\notag\\
	+ |\mathcal{H}_{\mathrm{S}}(f) - 1|^2~S_{\varphi}^{\mathrm{LO}}(f).
	\label{eqn:SpsiOLf}
\end{align}
The magnitude-squared transfer functions ${|\mathcal{H}_{\mathrm{S}}(f)|^2}$ and ${|\mathcal{H}_{\mathrm{S}}(f) - 1|^2}$ in Eq.~\eqref{eqn:SpsiOLf} depend on the type of sensor and are derived in the following.

\vspace{0.25cm}
\subsubsection{Resonator}
\label{subsubsec:phase_noise_phase_noise_in_open_loop_readout_system_resonator}

For a resonant sensor, the power spectral densities of the random phase fluctuations of the sensor and the amplifier are simply weighted by ${|\mathcal{H}_{\mathrm{S}}(f)|^2 = |\mathcal{H}_{\mathrm{R}}(f)|^2}$ (Eq.~\eqref{eqn:Hss} and Eq.~\eqref{eqn:HRfmagsq}). According to Eq.~\eqref{eqn:Hssminuseins} the transfer of the phase noise of the LO to the output of the system is given by
\begin{align}
	|\mathcal{H}_{\mathrm{R}}(f) - 1|^2 &= \frac{1}{1 + \left(\frac{f_{\mathrm{R}}}{2 Q f}\right)^2}.
	\label{eqn:HRfminuseinsmagsq}
\end{align}
Both phase noise transfer functions as a function of the frequency and for various quality factors are visualized in Fig.~\ref{fig:phase_noise_transfer_functions_resonator}. As expected, the phase noise of the sensor and the amplifier will be transformed unaltered to the open-loop system's output for frequencies inside the sensor's passband (green curves). The \SI{-3}{dB} cutoff frequency ${f_{\mathrm{L}} = f_{\mathrm{R}}/(2Q)}$ is called the \textit{Leeson frequency} \cite[p. 74]{Rubiola.2009} which is equal to half of the resonator's bandwidth $B_{\mathrm{R}}$. The phase noise of the oscillator is largely suppressed for low frequencies and low quality factors. However, both for increasing frequency and increasing quality factor the suppression decreases (blue curves). The reason is that the correlation of LO phase noise in both branches of the open-loop system decreases for higher frequencies and for longer relaxation time of the resonator.

\vspace{0.25cm}
\subsubsection{Delay Line}
\label{subsubsec:phase_noise_phase_noise_in_open_loop_readout_system_delay_line}

\begin{figure}[t]
	\centering
	\begin{subfigure}[t]{0.5\textwidth}
		\centering
		\includegraphics[width=1\linewidth]{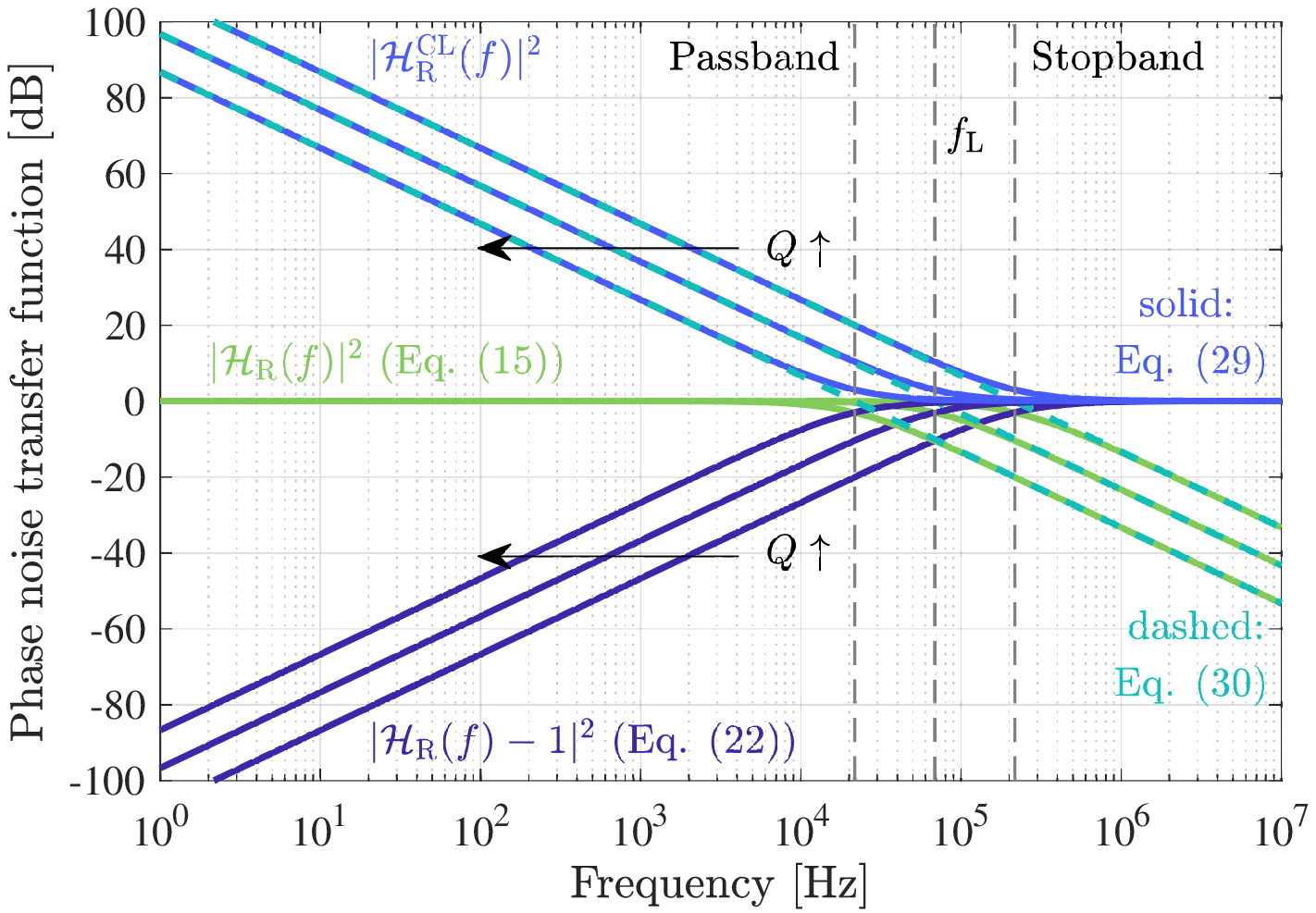}
		\caption{Resonant sensor}
		\label{fig:phase_noise_transfer_functions_resonator}
	\end{subfigure}
	~
	\begin{subfigure}[t]{0.5\textwidth}
		\centering
		\includegraphics[width=1\linewidth]{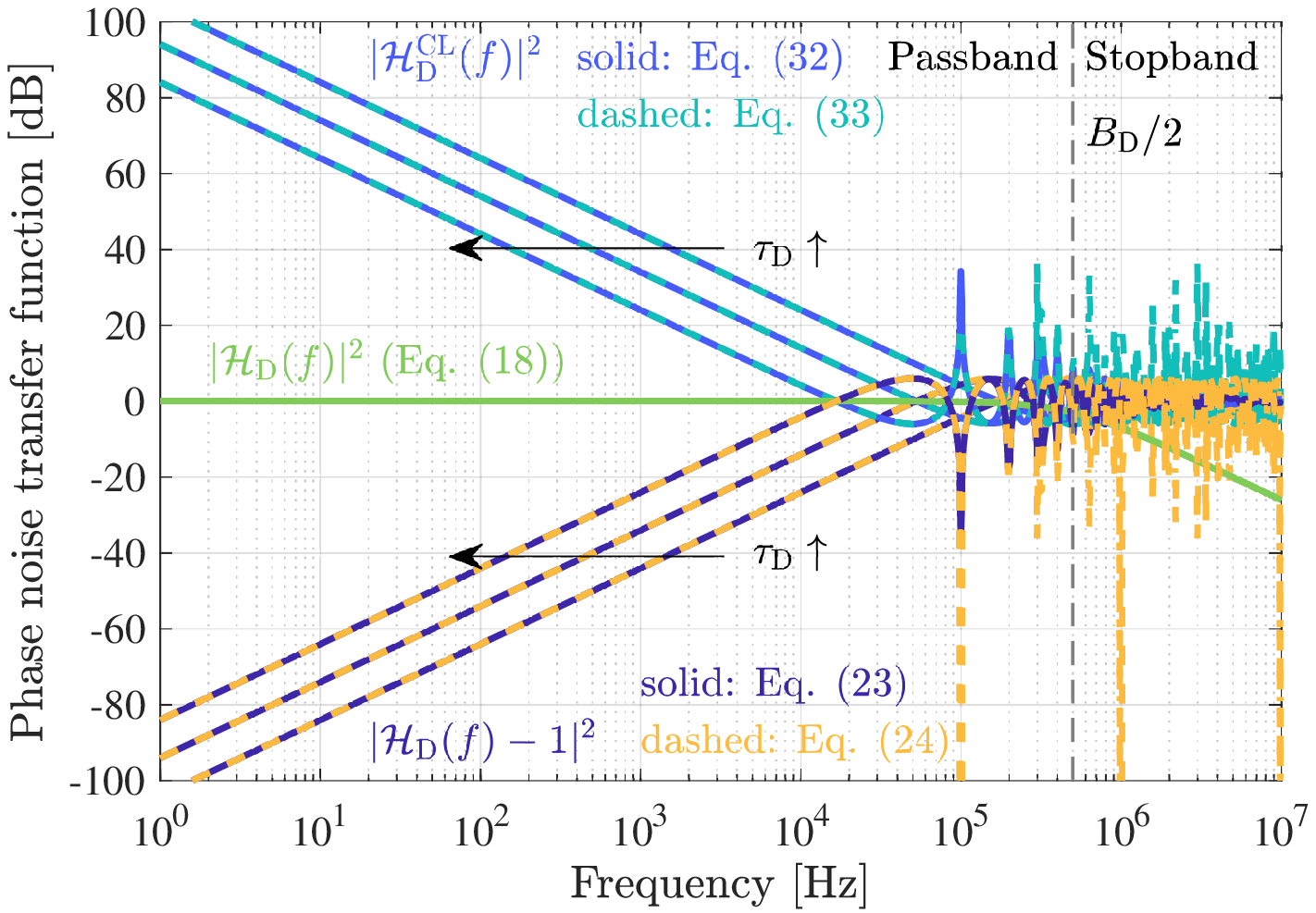}
		\caption{Delay line sensor}
		\label{fig:phase_noise_transfer_functions_delay_line}
	\end{subfigure}
	~
	\caption{Progression of the calculated phase noise transfer functions in open-loop and closed-loop sensor systems for various quality factors $Q$ of a resonant sensor (\subref{fig:phase_noise_transfer_functions_resonator}) and various delay times $\tau_{\mathrm{D}}$ of a delay line sensor (\subref{fig:phase_noise_transfer_functions_delay_line}). The chosen sensor parameters are ${f_{\mathrm{R}} = f_{\mathrm{D}} = \SI{434}{MHz}}$, ${Q = \{1000, 3162, 10000\}}$, ${B_{\mathrm{D}} = \SI{1}{MHz}}$, and ${\tau_{\mathrm{D}} = \{ \SI{1}{\mu s}, \SI{3.162}{\mu s}, \SI{10}{\mu s}\}}$.} 
	\label{fig:phase_noise_transfer_functions}
\end{figure}

For a delay line sensor in an open-loop system, the power spectral densities of the random phase fluctuations of the sensor and the amplifier are weighted by ${|\mathcal{H}_{\mathrm{S}}(f)|^2 = |\mathcal{H}_{\mathrm{D}}(f)|^2}$ (Eq.~\eqref{eqn:HDfmagsq}). According to Eq.~\eqref{eqn:Hssminuseins} the transfer of the phase noise of the LO to the output of the system is given by
\begin{align}
	|\mathcal{H}_{\mathrm{D}}(f) - 1|^2\notag\\
	&\hspace{-1.4cm}= \frac{2 + \left(\frac{2 f}{B_{\mathrm{D}}}\right)^2}{1 + \left(\frac{2 f}{B_{\mathrm{D}}}\right)^2} - \frac{2 \cos(2 \pi f \tau_{\mathrm{D}}) \sqrt{1 + \left(\frac{2 f}{B_{\mathrm{D}}}\right)^2}}{1 + \left(\frac{2 f}{B_{\mathrm{D}}}\right)^2}\label{eqn:HDfminuseinsmagsq}\\[0.2cm]
	&\hspace{-1.4cm} \approx 4 \sin^2(\pi f \tau_{\mathrm{D}}) \hspace{0.5cm} \text{for} \hspace{0.5cm} f \ll B_{\mathrm{D}}/2.\label{eqn:HDfminuseinsmagsqapprox}
\end{align}
The exact result in Eq.~\eqref{eqn:HDfminuseinsmagsq} takes into account the finite bandwidth of the sensor. For frequencies inside the sensor's passband (${f \ll B_{\mathrm{D}}/2}$) the expression distinctly simplifies and gives the same result calculated following another approach and verified by measurements in previous investigations \cite{Durdaut.2018}. All three phase noise transfer functions are depicted in Fig.~\ref{fig:phase_noise_transfer_functions_delay_line}. As for the previously discussed resonant sensor, the phase noise of the sensor and the amplifier will be transformed unaltered to the open-loop system's output for frequencies inside the sensor's passband (green curves), i.e. the \SI{-3}{dB} cutoff frequency ${B_{\mathrm{D}}/2}$. The phase noise of the oscillator (dark blue curved), again, is largely suppressed for low frequencies which, inside the sensor's passband, is well described by the approximation in Eq.~\eqref{eqn:HDfminuseinsmagsqapprox} (dashed light blue curves). Because the decreasing correlation of the LO phase noise in both branches of the open-loop system for higher delay times, the suppression decreases with $\tau_{\mathrm{D}}$.

\subsection{Phase Noise in Closed-Loop Readout System}
\label{subsec:phase_noise_phase_noise_in_closed_loop_readout_system}

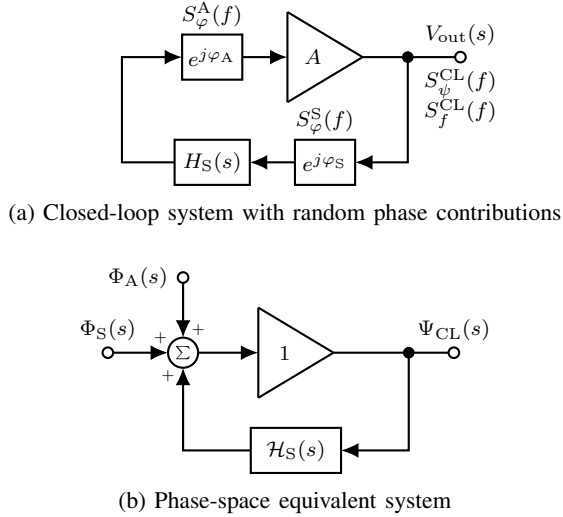
\begin{figure}[t]
	\centering
	\begin{subfigure}[t]{0.5\textwidth}
		\centering
		\trimbox{-0.75cm 0cm 0cm 0cm}{\usetikzlibrary{arrows.meta}

\begin{tikzpicture}

\draw[thick]  (-1.6,1.8) rectangle (-0.8,1.2);
\node  [align=center] at (-1.2,1.5) {\footnotesize $e^{j \varphi_\mathrm{A}}$};
\node  [align=center] at (-1.2,2.05) {\footnotesize $S_{\varphi}^{\mathrm{A}}(f)$};

\draw[thick,-{Latex[width=2mm]}] (-0.8,1.5) -- (-0.2,1.5);

\draw[thick] (-0.2,2.1) -- (-0.2,0.9);
\draw[thick] (0.8,1.5) -- (-0.2,2.1);
\draw[thick] (0.8,1.5) -- (-0.2,0.9);
\node  [align=center] at (0.14,1.5) {\footnotesize $A$};

\draw[thick,-{Latex[width=2mm]}] (0.8,1.5) -- (1.4,1.5) -- (1.4,0.1) -- (0.7,0.1);

\draw[thick]  (-0.1,0.4) rectangle (0.7,-0.2);
\node  [align=center] at (0.3,0.1) {\footnotesize $e^{j \varphi_\mathrm{S}}$};
\node  [align=center] at (0.3,0.65) {\footnotesize $S_{\varphi}^{\mathrm{S}}(f)$};

\draw[thick,-{Latex[width=2mm]}] (-0.1,0.1) -- (-0.7,0.1);

\draw[thick]  (-0.7,0.4) rectangle (-1.7,-0.2);
\node  [align=center] at (-1.2,0.1) {\footnotesize $H_{\mathrm{S}}(s)$};

\draw[thick,-{Latex[width=2mm]}] (-1.7,0.1) -- (-2.4,0.1) -- (-2.4,1.5) -- (-1.6,1.5);

\draw[fill=black] (1.4,1.5) circle (2pt);

\draw[thick,-] (1.4,1.5) -- (2.1,1.5);

\draw[thick,fill=white] (2.1,1.5) circle (2pt);
\node  [align=center] at (2.1,1.8) {\footnotesize $V_{\mathrm{out}}(s)$};

\node  [align=center] at (2.1,1.15) {\footnotesize $S_{\psi}^{\mathrm{CL}}(f)$};
\node  [align=center] at (2.1,0.75) {\footnotesize $S_{f}^{\mathrm{CL}}(f)$};

\end{tikzpicture}}
		\caption{Closed-loop system with random phase contributions}
		\label{fig:oscillator_rf}
	\end{subfigure}
	~
	\begin{subfigure}[t]{0.5\textwidth}
		\centering
		\trimbox{0cm 0cm 0cm -0.4cm}{\usetikzlibrary{arrows.meta}

\begin{tikzpicture}

\draw[thick] (-1.4,2.1) -- (-1.4,0.9);
\draw[thick] (-0.4,1.5) -- (-1.4,2.1);
\draw[thick] (-0.4,1.5) -- (-1.4,0.9);
\node  [align=center] at (-1.05,1.5) {\footnotesize $1$};

\draw[thick]  (-0.3,0.5) rectangle (-1.5,-0.1);
\node  [align=center] at (-0.9,0.2) {\footnotesize $\mathcal{H}_{\mathrm{S}}(s)$};

\draw[thick,-{Latex[width=2mm]}] (-0.4,1.5) -- (0.6,1.5) -- (0.6,0.2) -- (-0.3,0.2);
\draw[thick,-{Latex[width=2mm]}] (-1.5,0.2) -- (-2.4,0.2) -- (-2.4,1.3);
\node  [align=center] at (-2.6,1.2) {\tiny $+$};
\draw[thick,-{Latex[width=2mm]}] (-2.2,1.5) -- (-1.4,1.5);

\draw[thick,-{Latex[width=2mm]}] (-3.4,1.5) -- (-2.6,1.5);
\draw[thick,-{Latex[width=2mm]}] (-2.4,2.5) -- (-2.4,1.7);

\draw[thick,fill=white] (-2.4,1.5) circle (0.2cm);
\node  [align=center] at (-2.4,1.5) {\tiny $\sum$};

\draw[thick,fill=white] (-3.4,1.5) circle (2pt);
\draw[thick,fill=white] (-2.4,2.5) circle (2pt);

\node  [align=center] at (-3.4,1.8) {\footnotesize $\Phi_{\mathrm{S}}(s)$};
\node  [align=center] at (-2.7,1.7) {\tiny $+$};

\node  [align=center] at (-3,2.5) {\footnotesize $\Phi_{\mathrm{A}}(s)$};
\node  [align=center] at (-2.2,1.8) {\tiny $+$};

\draw[thick,-] (0.6,1.5) -- (1.2,1.5);
\draw[fill=black] (0.6,1.5) circle (2pt);
\draw[thick,fill=white] (1.2,1.5) circle (2pt);
\node  [align=center] at (1.2,1.8) {\footnotesize $\Psi_{\mathrm{CL}}(s)$};

\end{tikzpicture}}
		\caption{Phase-space equivalent system}
		\label{fig:oscillator_phase}
	\end{subfigure}
	~
	\caption{Basic structure of an closed-loop sensor readout system, i.e. an oscillator, together with the random phase contributions of the individual components (\subref{fig:oscillator_rf}). The use of the phase-space equivalent system (\subref{fig:oscillator_phase}) simplifies phase noise analysis as phase noise turns into additive noise but requires the determination of the phase-space equivalent transfer function of the sensor ${\mathcal{H}_{\mathrm{S}}(s)}$.}
	\label{fig:oscillator_rf_and_phase}
\end{figure}

Fig.~\ref{fig:oscillator_rf} depicts the closed-loop readout system, i.e. the oscillator, in the RF domain together with the random phase fluctuations of the sensor $\varphi_{\mathrm{S}}$ and the amplifier $\varphi_{\mathrm{A}}$. The related power spectral densities are denoted by ${S_{\varphi}^{\mathrm{S}}(f)}$ and ${S_{\varphi}^{\mathrm{A}}(f)}$. As described above, the phase-space representation of the system (Fig.~\ref{fig:oscillator_phase}) is more suited to calculate the overall phase noise at the output of the closed-loop system ${S_{\psi}^{\mathrm{CL}}(f)}$. Due to linearity, the Laplace transforms of the phase noise of the sensor ${\Phi_{\mathrm{S}}(s) = \mathcal{L}(\varphi_{\mathrm{S}})}$ and the amplifier ${\Phi_{\mathrm{A}}(s) = \mathcal{L}(\varphi_{\mathrm{A}})}$ can be arranged at any point inside the loop. Elementary feedback theory known from e.g. classical control theory or the analysis of operational amplifier circuits yields the phase noise transfer function of the closed-loop system
\begin{equation}
	\mathcal{H}_{\mathrm{CL}}(s) = \frac{\Psi_{\mathrm{CL}}(s)}{\Phi_{\mathrm{S}}(s)} = \frac{\Psi_{\mathrm{CL}}(s)}{\Phi_{\mathrm{A}}(s)} = \frac{1}{1-\mathcal{H}_{\mathrm{S}}(s)}.
	\label{eqn:HCLs}
\end{equation}
Thus, the overall power spectral density of the random phase fluctuations at the output of the oscillator as a function of both the phase noise of the sensor and the amplifier and the characteristic of the sensor yields
\begin{align}
	S_{\psi}^{\mathrm{CL}}(f) &= |\mathcal{H}_{\mathrm{CL}}(f)|^2~\left(S_{\varphi}^{\mathrm{S}}(f) + S_{\varphi}^{\mathrm{A}}(f)\right).
	\label{eqn:SpsiCLf}
\end{align}
According to the relation between random phase fluctuations and random frequency fluctuations in the time domain in Eq.~\eqref{eqn:fCLvonpsiCL}, the power spectral density of the random frequency fluctuations at the output of the oscillator in units of ${\mathrm{Hz}^2/\mathrm{Hz}}$ is given by
\begin{equation}
	S_{f}^{\mathrm{CL}}(f) = f^2 S_{\psi}^{\mathrm{CL}}(f).
	\label{eqn:SfCLf}
\end{equation}
In Eq.~\eqref{eqn:SpsiCLf} and also for Eq.~\eqref{eqn:SfCLf}, the magnitude-squared phase noise transfer function ${|\mathcal{H}_{\mathrm{CL}}(f)|^2}$ depends on the type of sensor and is derived in the following.

\vspace{0.25cm}
\subsubsection{Resonator}
\label{subsubsec:phase_noise_phase_noise_in_closed_loop_readout_system_resonator}

According to Eq.~\eqref{eqn:HCLs}, for a resonant sensor with the phase-space equivalent transfer function ${\mathcal{H}_{\mathrm{R}}(s)}$ from Eq.~\eqref{eqn:HRs} the phase noise transfer function of the closed-loop system yields
\begin{equation}
	\mathcal{H}_{\mathrm{R}}^{\mathrm{CL}}(s) = \frac{1}{1-\mathcal{H}_{\mathrm{R}}(s)} = 1 + \frac{\pi f_{\mathrm{R}}}{s Q}.
\end{equation}
Thus, the magnitude-squared phase noise transfer function that transforms the power spectral densities of the random phase fluctuations of the resonant sensor and the amplifier into oscillator phase noise (Eq.~\eqref{eqn:SpsiCLf}) results in
\begin{equation}
	|\mathcal{H}_{\mathrm{R}}^{\mathrm{CL}}(f)|^2 = 1 + \left( \frac{f_{\mathrm{R}}}{2 Q f} \right)^2.
\end{equation}
This equation is equal to the well-known \textit{Leeson formula} \cite{Leeson.1966} which simplifies to
\begin{equation}
	|\mathcal{H}_{\mathrm{R}}^{\mathrm{CL}}(f)|^2 \overset{f \ll f_{\mathrm{L}}}{\underset{}{\approx}} \left( \frac{f_{\mathrm{R}}}{2 Q f} \right)^2
	\label{eqn:HRCLfmagsq}
\end{equation}
for slow phase fluctuations below the \textit{Leeson frequency}. As it can be seen in Fig.~\ref{fig:phase_noise_transfer_functions_resonator} phase noise of the sensor and the amplifier is strongly raised in the closed-loop and even increases with the quality factor $Q$. This phenomenon is known as the \textit{Leeson effect}.

\vspace{0.25cm}
\subsubsection{Delay Line}
\label{subsubsec:phase_noise_phase_noise_in_closed_loop_readout_system_delay_line}

According to Eq.~\eqref{eqn:HCLs}, for a delay line sensor with the phase-space equivalent transfer function ${\mathcal{H}_{\mathrm{D}}(s)}$ from Eq.~\eqref{eqn:HDs} the phase noise transfer function of the closed-loop system yields
\begin{equation}
	\mathcal{H}_{\mathrm{D}}^{\mathrm{CL}}(s) = \frac{1}{1-\mathcal{H}_{\mathrm{D}}(s)} = \frac{1}{1 - \frac{e^{-s \tau_{\mathrm{D}}}}{\sqrt{1 + \left( \frac{2 f}{B_{\mathrm{D}}} \right)^2}}}.
\end{equation}
Thus, the magnitude-squared phase noise transfer function that transforms the power spectral densities of the random phase fluctuations of the delay line sensor and the amplifier into oscillator phase noise (Eq.~\eqref{eqn:SpsiCLf}) results in
\begin{align}
	|\mathcal{H}_{\mathrm{D}}^{\mathrm{CL}}(f)|^2\notag\\
	&\hspace{-1.0cm}= \frac{1 + \left(\frac{2 f}{B_{\mathrm{D}}}\right)^2}{\left(\frac{2 f}{B_{\mathrm{D}}}\right)^2 + 2 \left( 1 - \sqrt{1 + \left(\frac{2 f}{B_{\mathrm{D}}}\right)^2} \cos(2 \pi f \tau_{\mathrm{D}}) \right)}\\
	&\hspace{-1cm}\approx \frac{1}{2 \left( 1 - \cos(2 \pi f \tau_{\mathrm{D}}) \right)} \hspace{0.5cm} \text{for} \hspace{0.5cm} f \ll B_{\mathrm{D}}/2.\label{eqn:HDCLfmagsqapprox}
\end{align}
As for resonant sensors, in closed-loop systems the phase noise of the sensor and the amplifier is strongly raised and increases with the delay time $\tau_{\mathrm{D}}$ (Fig.~\ref{fig:phase_noise_transfer_functions_delay_line}).

\section{Limit of Detection}
\label{sec:limit_of_detection}
The frequency-dependent noise floor of a sensor system should always be given by a spectral density that is related to the unit of the physical quantity to be detected. For a physical quantity with the arbitrary unit $\mathrm{au}$ (see. Sec.~\ref{subsec:sensitivity_open_loop_and_closed_loop_readout_systems}) the representation of the sensor system's noise floor could be given as a power spectral density of the fluctuations of the arbitrary quantity in units of $\mathrm{au}^2/\mathrm{Hz}$. However, in general, it is more common to use the amplitude spectral density of the fluctuations of the arbitrary quantity in units of $\mathrm{au}/\sqrt{\mathrm{Hz}}$, referred to as limit of detection (LOD).

\subsection{Resonant Sensor}
\label{subsec:limit_of_detection_resonator}

With the expressions for the open-loop sensitivity and the power spectral density of random phase fluctuations from Eq.~\eqref{eqn:SPMresonator} and Eq.~\eqref{eqn:SpsiOLf}, respectively, the LOD in the open-loop system is defined by
\begin{align}
	\mathrm{LOD}_{\mathrm{R}}^{\mathrm{OL}}(f) &= \sqrt{\frac{S_{\psi}^{\mathrm{OL}}(f)}{\mathcal{S}_{\mathrm{PM}}^2}\label{eqn:LODROLf}}\\
	&\hspace{-1.8cm}= \frac{\sqrt{|\mathcal{H}_{\mathrm{R}}(f)|^2~\left(S_{\varphi}^{\mathrm{S}}(f) + S_{\varphi}^{\mathrm{A}}(f)\right) + |\mathcal{H}_{\mathrm{R}}(f) - 1|^2~S_{\varphi}^{\mathrm{LO}}(f)}}{2 \pi \tau_{\mathrm{R}} \mathcal{S}_{\mathrm{R}}}\notag.	
\end{align}
Due to the strong suppression of the local oscillator's phase noise by ${|\mathcal{H}_{\mathrm{R}}(f) - 1|^2}$ (Eq.~\eqref{eqn:HRfminuseinsmagsq}) it can be neglected such that the LOD yields
\begin{align}
	\mathrm{LOD}_{\mathrm{R}}^{\mathrm{OL}}(f) \approx \frac{|\mathcal{H}_{\mathrm{R}}(f)|~\sqrt{S_{\varphi}^{\mathrm{S}}(f) + S_{\varphi}^{\mathrm{A}}(f)}}{2 \pi \tau_{\mathrm{R}} \mathcal{S}_{\mathrm{R}}}.
	\label{eqn:LODROLfapprox1}
\end{align}
Referring to Fig.~\ref{fig:phase_noise_transfer_functions_resonator} and Eq.~\eqref{eqn:HRfmagsq} the LOD for a resonant sensor in an open-loop system further simplifies to
\begin{align}
	\mathrm{LOD}_{\mathrm{R}}^{\mathrm{OL}}(f) \approx \frac{f_{\mathrm{R}} \sqrt{S_{\varphi}^{\mathrm{S}}(f) + S_{\varphi}^{\mathrm{A}}(f)}}{2 Q \mathcal{S}_{\mathrm{R}}}
	\label{eqn:LODROLfapprox2}
\end{align}
 for frequencies ${f \ll f_{\mathrm{L}}}$ and relaxation time ${\tau_{\mathrm{R}} = Q/(\pi f_{\mathrm{R}})}$.

Based on the closed-loop sensitivity of a resonant sensor (Eq.~\eqref{eqn:SFMresonator}) and the expression for the power spectral density of the random frequency fluctuations from Eq.~\eqref{eqn:SfCLf} and Eq.~\eqref{eqn:SpsiCLf} the LOD in the closed-loop system is given by
\begin{align}
	\mathrm{LOD}_{\mathrm{R}}^{\mathrm{CL}}(f) &= \sqrt{\frac{S_{f}^{\mathrm{CL}}(f)}{\mathcal{S}_{\mathrm{FM}}^2}} = \frac{f~\sqrt{S_{\psi}^{\mathrm{CL}}(f)}}{\mathcal{S}_{\mathrm{R}}}\notag\\
	&= \frac{f~|\mathcal{H}_{\mathrm{R}}^{\mathrm{CL}}(f)|~\sqrt{S_{\varphi}^{\mathrm{S}}(f) + S_{\varphi}^{\mathrm{A}}(f)}}{\mathcal{S}_{\mathrm{R}}}.
\end{align}
With Eq.~\eqref{eqn:HRCLfmagsq} the approximated LOD for a resonant sensor operated in its passband and in a closed-loop system results in
\begin{align}
	\mathrm{LOD}_{\mathrm{R}}^{\mathrm{CL}}(f) &\approx \frac{f \left( \frac{f_{\mathrm{R}}}{2 Q f} \right) \sqrt{S_{\varphi}^{\mathrm{S}}(f) + S_{\varphi}^{\mathrm{A}}(f)}}{\mathcal{S}_{\mathrm{R}}}\\
	&= \frac{f_{\mathrm{R}}~\sqrt{S_{\varphi}^{\mathrm{S}}(f) + S_{\varphi}^{\mathrm{A}}(f)}}{2 Q \mathcal{S}_{\mathrm{R}}} = \mathrm{LOD}_{\mathrm{R}}^{\mathrm{OL}}(f)
\end{align}
which is equal to the LOD for a resonant sensor in an open-loop readout system described by Eq.~\eqref{eqn:LODROLfapprox2}.

\subsection{Delay Line Sensor}
\label{subsec:limit_of_detection_delay_line}

With the expressions for the open-loop sensitivity and the power spectral density of random phase fluctuations from Eq.~\eqref{eqn:SPMdelayline} and Eq.~\eqref{eqn:SpsiOLf}, respectively, the LOD in the open-loop system is defined by
\begin{align}
	\mathrm{LOD}_{\mathrm{D}}^{\mathrm{OL}}(f) &= \sqrt{\frac{S_{\psi}^{\mathrm{OL}}(f)}{\mathcal{S}_{\mathrm{PM}}^2}}\label{eqn:LODDOLf}\\
	&\hspace{-1.8cm}= \frac{\sqrt{|\mathcal{H}_{\mathrm{D}}(f)|^2~\left(S_{\varphi}^{\mathrm{S}}(f) + S_{\varphi}^{\mathrm{A}}(f)\right) + |\mathcal{H}_{\mathrm{D}}(f) - 1|^2~S_{\varphi}^{\mathrm{LO}}(f)}}{\mathcal{S}_{\mathrm{D}}}\notag.	
\end{align}
Again, due to the strong suppression of the local oscillator's phase noise by ${|\mathcal{H}_{\mathrm{D}}(f) - 1|^2}$ (Eq.~\eqref{eqn:HDfminuseinsmagsqapprox}) it can be neglected such that the LOD yields
\begin{align}
	\mathrm{LOD}_{\mathrm{D}}^{\mathrm{OL}}(f) \approx \frac{|\mathcal{H}_{\mathrm{D}}(f)|~\sqrt{S_{\varphi}^{\mathrm{S}}(f) + S_{\varphi}^{\mathrm{A}}(f)}}{\mathcal{S}_{\mathrm{D}}}.
	\label{eqn:LODDOLfapprox1}
\end{align}
Referring to Fig.~\ref{fig:phase_noise_transfer_functions_delay_line} and Eq.~\eqref{eqn:HDfmagsq} the LOD for a delay line sensor in an open-loop system reduces to
\begin{align}
	\mathrm{LOD}_{\mathrm{D}}^{\mathrm{OL}}(f) \approx \frac{\sqrt{S_{\varphi}^{\mathrm{S}}(f) + S_{\varphi}^{\mathrm{A}}(f)}}{\mathcal{S}_{\mathrm{D}}}
	\label{eqn:LODDOLfapprox2}
\end{align}
for frequencies ${f \ll B_{\mathrm{D}}/2}$.

Considering the closed-loop sensitivity of a delay line sensor (Eq.~\eqref{eqn:SFMdelayline}) and the expression for the power spectral density of the random frequency fluctuations from Eq.~\eqref{eqn:SfCLf} and Eq.~\eqref{eqn:SpsiCLf} the LOD in the closed-loop system is given by
\begin{align}
	\mathrm{LOD}_{\mathrm{D}}^{\mathrm{CL}}(f) &= \sqrt{\frac{S_{f}^{\mathrm{CL}}(f)}{\mathcal{S}_{\mathrm{FM}}^2}} = \frac{f~\sqrt{S_{\psi}^{\mathrm{CL}}(f)}}{\frac{\mathcal{S}_{\mathrm{D}}}{2 \pi \tau_{\mathrm{D}}}}\notag\\
	&= \frac{f~|\mathcal{H}_{\mathrm{D}}^{\mathrm{CL}}(f)|~\sqrt{S_{\varphi}^{\mathrm{S}}(f) + S_{\varphi}^{\mathrm{A}}(f)}}{\frac{\mathcal{S}_{\mathrm{D}}}{2 \pi \tau_{\mathrm{D}}}}.
\end{align}
With Eq.~\eqref{eqn:HDCLfmagsqapprox} the approximated LOD for a delay line sensor operated in its passband results in
\begin{align}
	\mathrm{LOD}_{\mathrm{D}}^{\mathrm{CL}}(f) &\approx \frac{2 \pi \tau_{\mathrm{D}}~f~\sqrt{S_{\varphi}^{\mathrm{S}}(f) + S_{\varphi}^{\mathrm{A}}(f)}}{\mathcal{S}_{\mathrm{D}} \cdot \sqrt{2 \left(1-\cos(2 \pi f \tau_{\mathrm{D}})\right)}}.
	\label{eqn:LODDCLf}	
\end{align}
Using the \textit{Taylor series approximation} ${\cos(a) \approx 1 - a^2/2}$ (valid for ${f \tau_{\mathrm{D}} \ll 0.1}$), the LOD for a delay line sensor in a closed-loop system reduces to
\begin{align}
	\mathrm{LOD}_{\mathrm{D}}^{\mathrm{CL}}(f) &\approx \frac{\sqrt{S_{\varphi}^{\mathrm{S}}(f) + S_{\varphi}^{\mathrm{A}}(f)}}{\mathcal{S}_{\mathrm{D}}} = \mathrm{LOD}_{\mathrm{D}}^{\mathrm{OL}}(f)
\end{align}
which is equal to the LOD for a delay line sensor in an open-loop readout system described by Eq.~\eqref{eqn:LODDOLfapprox2}.

\begin{figure}[t]
	\centering
	\includegraphics[width=0.5\textwidth]{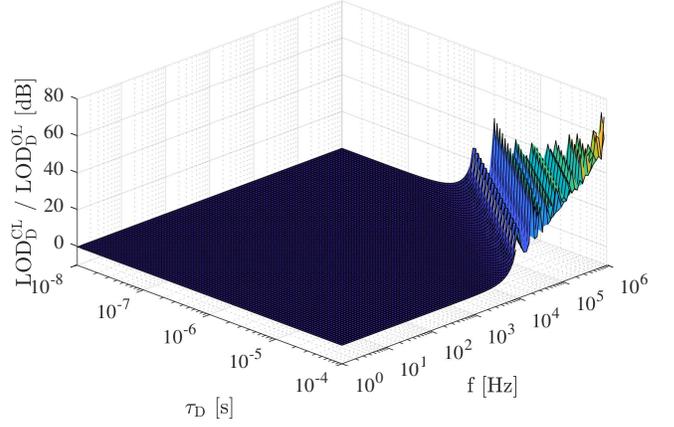}
	\caption{Ratio between the LOD in a closed-loop system (Eq.~\eqref{eqn:LODDCLf}) an for the LOD in an open-loop system (Eq.~\eqref{eqn:LODDOLfapprox2}) for delay line sensors. Both systems are equal for a wide range of frequencies and delay times, whereas the open-loop system is superior for sensors with large delay times and for the detection of very fast processes.}
	\label{fig:LOD_D_CL_vs_LOD_D_OL_for_BD_to_infty}
\end{figure}

As already mentioned, this equivalence relies on the \textit{Taylor series approximation} which is only valid for ${f \tau_{\mathrm{D}} \ll 0.1}$, thus for low frequencies $f$ and small delay times $\tau_{\mathrm{D}}$. The ratio between the LOD in an open-loop system (Eq.~\eqref{eqn:LODDOLfapprox2}) and the LOD in a closed-loop system (Eq.~\eqref{eqn:LODDCLf}) is depicted in Fig.~\ref{fig:LOD_D_CL_vs_LOD_D_OL_for_BD_to_infty} and confirms the equality of the two readout systems for a wide range of frequencies and delay times. However, because a delay line oscillator exhibits a distinct increase of phase noise at frequencies ${f = n/\tau_{\mathrm{D}}},~n \in \mathbb{N}^{+}$ \cite[p. 142 f.]{Rubiola.2009} the LOD in an open-loop system is superior when sensors with large delay times are used or when the physical quantity to detect changes very fast.

\section{Time Domain Uncertainty}
\label{sec:time_domain_uncertainty}
\begin{table}[!t]
\renewcommand{\arraystretch}{1.5}
\caption{Relations between the definitions of various noise processes and drift and the expressions for the Allan variance.}
\label{tab:noise_processes_and_avar}
\centering
\begin{tabular}{|c|c|c|}
\hline
Process				& Definition 																		& AVAR \\ \hline
White noise   & $S_y(f) = h_{0} $ 														& $\sigma_{y}^2(\tau) = \frac{h_{0}}{2 \tau}$ \\ \hline
Flicker noise & $S_y(f) = \frac{h_{-1}}{f} $ 									& $\sigma_{y}^2(\tau) = 2 \ln(2) h_{-1}$ \\ \hline
Random walk   & $S_y(f) = \frac{h_{-2}}{f^2} $ 								& $\sigma_{y}^2(\tau) = \frac{4 \pi^2}{6} h_{-2} \tau$ \\ \hline
Linear drift  & $D_y = \frac{\mathrm{d}y(t)}{\mathrm{d}t} $   & $\sigma_{y}^2(\tau) = \frac{1}{2} D_y^2 \tau^2$ \\ \hline
\end{tabular}
\end{table}

\begin{figure}[t]
	\centering
	\trimbox{0cm 0cm 0cm 0cm}{\usetikzlibrary{arrows.meta}

\begin{tikzpicture}


\draw[thick,-{Latex[width=2mm]}] (0,0) -- (7.2,0);
\draw[thick,-{Latex[width=2mm]}] (0,0) -- (0,2.7);
\node  [align=center] at (0,2.95) {$\sigma_y^2(\tau)$};
\node  [align=center] at (7.45,0) {$\tau$};

\draw[thick] (0.2,1.1) -- (1.7,0.4) -- (4,0.4) -- (5.7,1) -- (6.8,2.4);
\node  [align=center, rotate=-25] at (0.9,0.62) {\footnotesize white noise};
\node  [align=center, rotate=-25] at (0.99,1.07) {\footnotesize $\frac{h_0}{2 \tau}$};
\node  [align=center, rotate=0] at (2.85,0.24) {\footnotesize flicker noise};
\node  [align=center, rotate=0] at (2.85,0.6) {\footnotesize $2 \ln(2) h_{-1}$};
\node  [align=center, rotate=20] at (4.84,0.53) {\footnotesize random walk};
\node  [align=center, rotate=20] at (4.82,0.98) {\footnotesize $\frac{4 \pi^2}{6} h_{-2} \tau$};
\node  [align=center, rotate=51] at (6.37,1.59) {\footnotesize linear drift};
\node  [align=center, rotate=51] at (6.16,2.01) {\footnotesize $\frac{1}{2} D_y^2 \tau^2$};

\draw[thick,densely dotted] (1.7,0.8) -- (1.7,-0.2);
\node  [align=center] at (1.7,-0.55) {\footnotesize $\frac{1}{4 \ln(2)} \frac{h_0}{h_{-1}}$};

\draw[thick,densely dotted] (4,0.8) -- (4,-0.2);
\node  [align=center] at (4,-0.55) {\footnotesize $\frac{3 \ln(2)}{\pi^2} \frac{h_{-1}}{h_{-2}}$};

\draw[thick,densely dotted] (5.7,1.4) -- (5.7,-0.2);
\node  [align=center] at (5.7,-0.55) {\footnotesize $\frac{4 \pi^2}{3} \frac{h_{-2}}{D_y^2}$};

\end{tikzpicture}}
	\caption{Schematic progression of the Allan variance $\sigma_{y}^2(\tau)$ as a function of the measurement time $\tau$ for various noise processes and linear drift.}
	\label{fig:avar}
\end{figure}
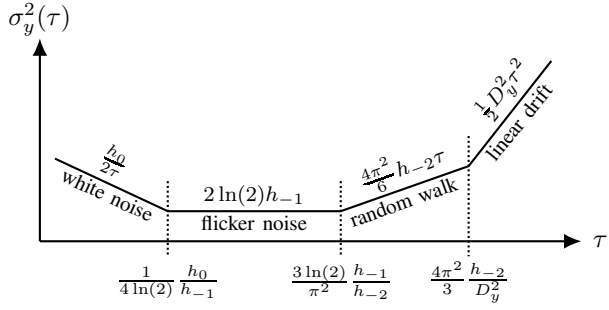

The sensor system's output is most often exploited as a continuous stream of values 
\begin{align}
	\overline{y}_k = \frac{1}{\tau} \int_{k\tau}^{(k+1)\tau}{y(t)~\mathrm{d}t},
	\label{eqn:yk}
\end{align}
each averaged over a suitable time $\tau$ with ${k \in \mathbb{N}_0}$ (please do not confuse $\tau$ with the delay time of a delay line sensor $\tau_{\mathrm{D}}$ or the relaxation time of a resonant sensor $\tau_{\mathrm{R}}$). It is therefore appropriate to describe the sensor system's noise in terms of a two-sample variance, also called \textit{Allan variance} (AVAR) \cite{Barnes.1971,Stein.2009} which is defined as 
\begin{align}
	\sigma_{y}^2(\tau) = \frac{1}{2} \mathbb{E} \left\{ \left[ \overline{y}_{k+1} - \overline{y}_k \right]^2 \right\}
	\label{eqn:sigmaysqtau}
\end{align}
where $\overline{y}_{k+1}$ and $\overline{y}_k$ are two values of $y(t)$ averaged on contiguous time slots of duration $\tau$ and $\mathbb{E}\{ \cdot \}$ denotes the mathematical expectation operator. Using a weighted average in Eq.~\eqref{eqn:yk} results in other types of variances, like the modified Allan Variance \cite{Allan.1981,Rubiola.2005}, the Parabolic variance \cite{Vernotte.2016,Benkler.2015}, etc., which is less common for sensors. Traditionally, $y(t)$ is the fractional frequency fluctuation ${y(t) = (\Delta f_0)(t)/f_0}$. However, the AVAR is a general tool and $y(t)$ can be replaced with any quantity, either absolute or fractional. In all experiments, the expectation $\mathbb{E}\{ \cdot \}$ is replaced with the average on a suitable number of realizations. The Allan variance can be seen as an extension of the classical variance, where the lowpass effect resulting from the difference ${\overline{y}_{k+1} - \overline{y}_k}$ provides the additional property that the AVAR converges for flicker and random walk processes, and even for a linear drift. These processes are of great interest for oscillators and sensor systems. Interestingly, random walk and drift in electronics are sometimes misunderstood, and both described with a single parameter called \textit{aging} (see for example \cite{Rubiola.2001}). The quantity $\sigma_{y}(\tau)$ is the statistical uncertainty, also referred to as \textit{Allan deviation} (ADEV), which depends on the measurement time $\tau$ (Fig.~\ref{fig:avar}) and can be calculated from the power spectral density $S_y(f)$ of random fluctuations of $y$ (Tab.~\ref{tab:noise_processes_and_avar}). The uncertainty decreases proportionally with ${1/\sqrt{\tau}}$ for white noise processes and attains its minimum in the flicker region ${( \tau_1 \approx 0.36~h_0/h_{-1} < \tau < \tau_2 \approx 0.21~h_{-1}/h_{-2} )}$ where the uncertainty is independent of $\tau$. This identifies $\tau_1$ as the optimum measurement time, to the extent that the lowest uncertainty is achieved in the shortest measurement time. Beyond $\tau_2$, the uncertainty degrades. 

\begin{table}[!t]
\renewcommand{\arraystretch}{1.5}
\caption{Coefficients describing white phase noise ($h_{0}$) and flicker phase noise ($h_{-1}$) at the output of an open-loop system. The white phase noise and flicker phase noise of the sensor is described by $b_{0}$ and $b_{-1}$, respectively.}
\label{tab:coeff_h_open_loop}
\centering
\begin{tabular}{c|c|c|c|c|}
																																																													\cline{2-3}
																& \multicolumn{2}{c|}{Open-loop system} 																							 \\ \cline{2-3} 
																& Resonator     														& Delay line     													 \\ \hline
\multicolumn{1}{|c|}{$h_{-1}$} 	& $|\mathcal{H}_{\mathrm{R}}(f)|^2 b_{-1}$  & $|\mathcal{H}_{\mathrm{D}}(f)|^2 b_{-1}$ \\ \hline
\multicolumn{1}{|c|}{$h_{0}$}  	& $|\mathcal{H}_{\mathrm{R}}(f)|^2 b_{0}$   & $|\mathcal{H}_{\mathrm{D}}(f)|^2 b_{0}$  \\ \hline
\end{tabular}
\end{table}

\begin{table}[!t]
\renewcommand{\arraystretch}{1.5}
\caption{Coefficients describing white frequency noise ($h_{0}$) and flicker frequency noise ($h_{-1}$) at the output of a closed-loop system. The white phase noise and flicker phase noise of the sensor is described by $b_{0}$ and $b_{-1}$, respectively.}
\label{tab:coeff_h_closed_loop}
\centering
\begin{tabular}{c|c|c|c|c|}
																																																																															 \cline{2-3}
																& \multicolumn{2}{c|}{Closed-loop system} 																																									\\ \cline{2-3} 
																& Resonator      																							& Delay line      																						\\ \hline
\multicolumn{1}{|c|}{$h_{-1}$} 	& $|\mathcal{H}_{\mathrm{R}}^{\mathrm{CL}}(f)|^2 f^2 b_{-1}$ 	& $|\mathcal{H}_{\mathrm{D}}^{\mathrm{CL}}(f)|^2 f^2 b_{-1}$  \\ \hline
\multicolumn{1}{|c|}{$h_{0}$}  	& $|\mathcal{H}_{\mathrm{R}}^{\mathrm{CL}}(f)|^2 f^2 b_{0}$  	& $|\mathcal{H}_{\mathrm{D}}^{\mathrm{CL}}(f)|^2 f^2 b_{0}$   \\ \hline
\end{tabular}
\end{table}

At the output of the sensor system the quantity of interest is represented by a \textit{phase} in case of the open-loop system, and represented by a \textit{frequency} in case of the closed-loop system, i.e. the oscillator. Consequently, the optimum measurement time $\tau_1$ is given by the intercept point between white phase noise (${i = 0}$) and flicker phase noise (${i = -1}$) for the open-loop system, and by the intercept point between white frequency noise (${i = -2}$) and flicker frequency noise (${i = -3}$) for the closed-loop system, respectively. Following the expressions for the power spectral densities of random phase fluctuations $S_{\psi}^{\mathrm{OL}}(f)$ (Eq.~\eqref{eqn:SpsiOLf}), and random frequency fluctuations $S_{f}^{\mathrm{CL}}(f)$ (Eq.~\eqref{eqn:SfCLf}), the coefficients $h_{-1}$ and $h_{0}$ result in expressions as listed in Tab.~\ref{tab:coeff_h_open_loop} and Tab.~\ref{tab:coeff_h_closed_loop} when considering only the phase noise of the sensor as ${S_{\varphi}^{\mathrm{S}}(f) = b_{-1}/f + b_0}$. Thus, the optimum measurement time
\begin{align}
	\tau_1 = \frac{1}{4 \ln(2)} \frac{h_0}{h_{-1}} = \frac{1}{4 \ln(2)} \frac{b_0}{b_{-1}} \approx 0.36 \frac{b_0}{b_{-1}}
\end{align}
turns out to be the same for open-loop and closed-loop systems as well as for both types of sensors. Our conclusion, that the two measurement methods are equivalent, relates to the sensor systems only, assuming that these are ideal. However, the shown derivations can be easily extended for ${S_{\varphi}^{\mathrm{A}}(f) \neq 0}$ and ${S_{\varphi}^{\mathrm{LO}}(f) \neq 0}$, at least numerically. It turns out that the background noise of a phase detector (used for the differential phase measurement in the open-loop system) is lower than the background of a frequency detector, i.e. a frequency counter. The reason is that the phase meter is a dedicated device, specialized for the phase detection in a narrow range around a given frequency. Overall, this kind of measurement relies on the principle of a lock-in amplifier, whose bandwidth is determined by a lowpass filter. By contrast, a frequency counter is a general-purpose device suitable for a wide range of input frequencies. Consequently, the statistical uncertainty is affected by the wide noise bandwidth.

\section{Conclusion}
\label{sec:conclusion}
In this paper, phase noise in open-loop and closed-loop readout systems for resonant surface acoustic wave (SAW) sensors and SAW delay line sensors is investigated. Comprehensive derivations are presented which analytically describe the phase noise in the various sensor systems. Based on these results and together with the sensitivities of the sensors in both systems, equivalence in terms of the minimum achievable limit of detection and the optimum measurement time between open-loop and closed-loop operation is shown for both types of sensors. Thus, the mode of operation should be chosen based on the availability of the needed low-noise electronic components and the complexity of the resulting system. For both readout structures, the random phase fluctuations introduced by the preamplifier directly add up with the sensor-intrinsic phase noise which is why the amplifier always needs to be chosen very carefully. As opposed to this, phase noise of the local oscillator in open-loop systems is usually largely suppressed. The presented results are not only valid for SAW devices but are also applicable to all kinds of phase sensitive sensors.

\section*{Acknowledgment}
This work was supported (1) by the German Research Foundation (Deutsche Forschungsgemeinschaft, DFG) through the Collaborative Research Centre CRC 1261 \textit{Magnetoelectric Sensors: From Composite Materials to Biomagnetic Diagnostics}, (2) by the ANR Programme d'Investissement d'Avenir (PIA) under the Oscillator IMP project and the FIRST-TF network, and (3) by grants from the R\'{e}gion Bourgogne Franche-Comt\'{e} intended to support the PIA.

\ifCLASSOPTIONcaptionsoff
  \newpage
\fi




\bibliographystyle{IEEEtran}
\bibliography{mybibfile}


%








\end{document}